\def\BJ{B$_{\rm J}$}
\def\RF{R$_{\rm F}$}
\def\AA{\hbox{${\rm A}\hskip-0.53em^{^{\circ}}$}}

\def\mag{\hbox{$\;.\!\!\!^m$}}

\def\muspc{\hskip 0.15 em}

\hyphenpenalty=50

\input aa.cmmb
\input psfig
\pageno=1
\MAINTITLE{Field \#3 of the Palomar-Groningen Survey}
\SUBTITLE{I. Variable stars at the edge of the Sagittarius dwarf galaxy
\FOOTNOTE{Table 1 and the finding charts of the variable stars 
are only available in electronic form at the CDS 
via anonymous {\tt ftp} to {\tt cdsarc.u-strasbg.fr (130.79.128.5)}}}
\AUTHOR{Y.K. Ng@1 and M. Schultheis@2}
\INSTITUTE{
 @1
 IAP, CNRS, 98bis Boulevard Arago, F-75014 Paris, France 
 ({\tt ng{\char64}iap.fr}) 

 @2
 Institut f\"ur Astronomie der Universit\"at Wien,
 T\"urkenschanzstra{\ss}e 17, A-1180 Wien, Austria
 \newline
 \quad({\tt schultheis{\char64}astro.ast.univie.ac.at})}
\DATE{Received 12 August 1996 / Accepted 19 September 1996}

\ABSTRACT{A catalogue is presented with variable
(RR Lyrae, semiregular and Mira) stars
located inside field \#3 of the Palomar-Groningen Survey
at the outer edge of the Sagittarius dwarf galaxy.
One of the semiregular variables is a carbon star, comparable with those
found by Azzopardi et al. (1991). Serendipity provides the 
suggestion, that their
carbon stars might not be located inside, but behind the bulge 
in the Sagittarius dwarf galaxy.
}

\THESAURUS{(08.03.1, 08.22.3, 11.12.1)}

\maketitle
\MAINTITLE{Field \#3 of the Palomar-Groningen Survey I.}
\SUBTITLE{Variable stars at the edge of the Sagittarius dwarf galaxy}

\titlea{Introduction}
The Palomar-Groningen field \#3 (hereafter referred to as PG3) 
was searched for variable stars by Plaut (1971). 
The variables discovered by Plaut were re-examined by Wesselink 
(1987), using UKST \BJ\ and \RF\ Schmidt plates. 
A new blink done on 10\% of the total PG3 area
on the original 103aO plate material used by 
Plaut, resulted in the discovery of some additional variables.
It demonstrated that the completeness of the variable star
catalogue is better for variables with larger amplitudes. 
The variables from this catalogue were subject in various 
studies: the RR Lyrae (Oort and Plaut 1975, Wesselink 1987),
the Mira variables (Blommaert 1992), and the 
semiregular variables (Schultheis et al. 1996, hereafter
referred to as Paper II).
The non-variable stars in PG3 were studied by Ng (1994)
and Ng et al. (1995).
\par
The Sagittarius dwarf galaxy was first identified by 
Ibata, Gilmore \& Irwin (1994, hereafter referred to as IGI). 
Alard (1996) and Mateo et al. (1996)
found that this galaxy is spatially more extended then found by IGI.
Some stars from the Sagittarius dwarf galaxy could be
present in the the PG3 variable star catalog from Plaut (1971) and 
Wesselink (1987). \hfill\break
In this paper we present a small catalogue
of PG3 variable stars, which are likely member of the 
Sagittarius dwarf galaxy. The selection criteria and the catalogue
are presented in Sect.~2. A discussion is given in Sect.~3, 
about the implications, that the carbon star in this sample
has, on those found in the direction of the galactic bulge
by Azzopardi et~al. (1991).

\titlea{Selection and catalogue}
Alard (1996) demonstrated, that the faint peak in 
a bimodal distribution of the RR Lyrae stars 
of Bailey type {\it ab} is due to a contribution from  
the Sagittarius dwarf galaxy. Application of this method 
to the PG3 RR Lyrae stars gives 12 variables of Bailey type {\it ab}
(hereafter referred to as RR{\it ab}\/ and RR{\it c}\/ 
for Bailey type {\it c})
and 4 RR{\it c}\/ stars. Figure~1 shows the distance modulus distribution 
from which the RR{\it ab}\/ stars were selected.
Following Wesselink (1987) we adopted for the absolute magnitude of
the RR{\it ab}: \hbox{M$_{\rm B_{\rm J}}$\muspc=\muspc0\mag79}
and \hbox{M$_{\rm R_{\rm F}}$\muspc=\muspc0\mag45}.
The difference in the distribution between the two passbands is 
due to extinction. From Fig.~1 we obtain for PG3 a mean colour excess 
\hbox{E(\BJ--\RF)\muspc=\muspc0\mag20\muspc$\pm$\muspc0\mag05}.
The stars with distance modulus larger than 17\mag5 are selected as possible
member of the Sagittarius dwarf galaxy. The RR{\it c}\/ candidates are 
selected in a similar way, straight from their magnitude distribution.
\hfill\break
This method cannot be applied to select the long period variables
in the visual passbands, because of their large amplitudes.
In the near-infrared passbands the amplitude of their
variations is significantly smaller.
From the K-magnitude distribution  (Paper~II, Fig.~2) 1~Mira and 
5~semiregular variables are thus selected.
The actual number of stars might be even larger, because we do
not have near-IR photometry for all PG3 long period variables.
Table~1 lists the whole catalogue of PG3 variables which are possibly
located in the Sagittarius dwarf galaxy. The finding charts 
($2\farcm67\times2\farcm67$; north is top and east is left) are only 
available in electronic form at the CDS. The identification 
of the stars was made by Wesselink (1987).

\par
\begfigps 5.5cm
\psfig{file=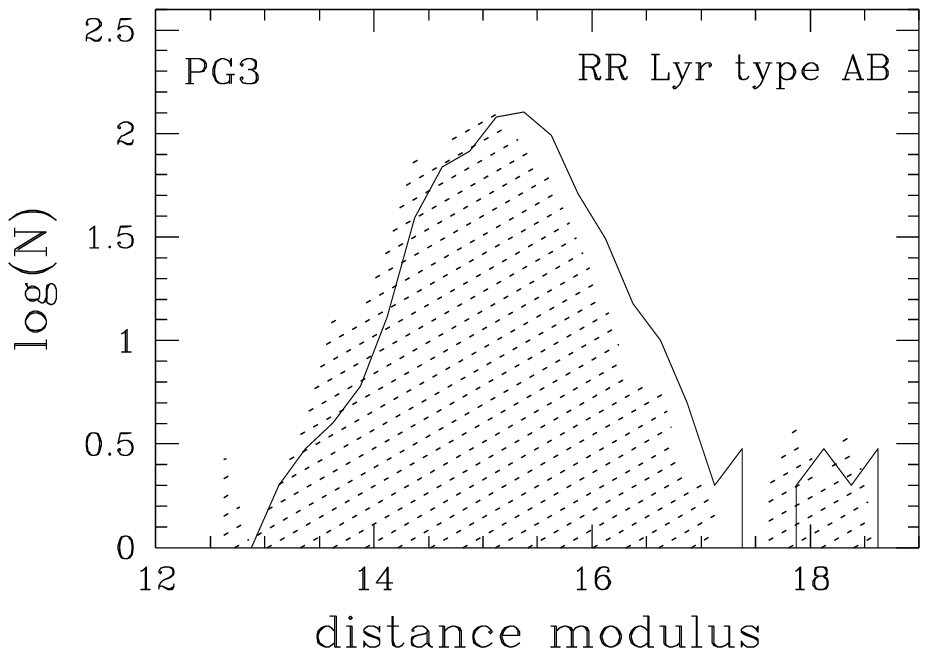,height=5.5cm,width=8.6cm}
\figure{1}{Distance modulus distribution of the 
RR Lyrae stars Bailey type {\it ab}
in PG3 (data from Wesselink 1987, see text for additional details). 
The solid line shows the distribution obtained from the \BJ-magnitudes
and the dotted distribution is obtained from the \RF-magnitudes.}
\endfig

\titlea{Discussion}
The sample is too small to estimate reliably from the RR Lyrae 
the distance to 
the Sagittarius dwarf galaxy. Furthermore, we are looking at the edge
of the dwarf galaxy and the RR Lyrae sample might be 
a mixture of stars from the galactic halo and the dwarf galaxy.
An indication that we are possible 
dealing with a mixed sample can be obtained as follows.
Select from the sample those RR{\it ab} stars for 
which extinction corrections are possible from their colours
at minimum light (i.e. \hbox{Q\muspc=\muspc0}:
both the period and the classification are correct). 
There are only four of those stars in the sample. 
If the extinction correction is done as described by 
Wesselink (1987) and Alard (1996) we have to discard two 
stars from our consideration, because one 
is too blue (\#1371) and another one is probably too 
faint (\#1524). For the two remaining stars we apply the
mean reddening correction mentioned in Sect.~2.
This gives a mean distance modulus for these stars of
\hbox{({\it m}--M)$_0$\muspc=\muspc17\mag05\muspc$\pm$\muspc0\mag15}
or a distance of \hbox{26\muspc$\pm$\muspc2~kpc}.
The distance modulus is in good agreement with 
\hbox{({\it m}--M)$_0$\muspc=\muspc17\mag02\muspc$\pm$\muspc0\mag19}
obtained by Mateo et al. (1995). The distance
is within the uncertainties of \hbox{24\muspc$\pm$\muspc2~kpc},
obtained by Alard (1996) from a much larger sample of stars.
The mean distance of the RR{\it ab}\/ stars in Table~1 is about 32~kpc. 
This might be due to a contribution of stars from the galactic halo 
or it might be due to an extension along the line of sight of the dwarf galaxy.
Radial velocities of the stars might help to distinguish the two
scenarios from each other. 
\par
\begfigps 5.5cm
\psfig{file=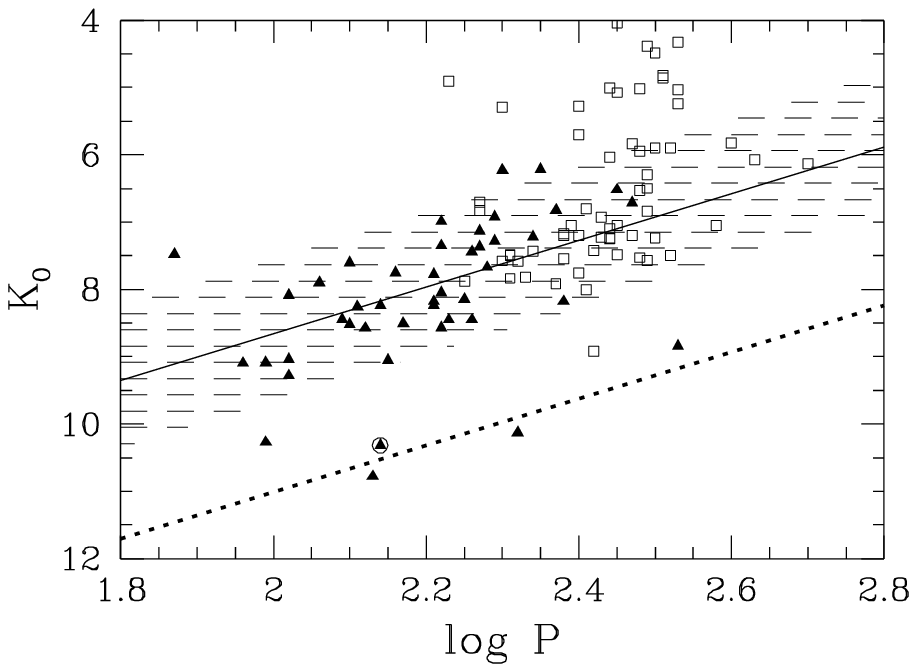,height=5.5cm,width=8.6cm}
\figure{2}{Period-K$_0$ relation for Mira (open square) and 
semiregular (triangle) variables in PG3. 
The open circle indicates a carbon star among these variables.
The thick solid line is the relation
obtained by Glass et al. (1995) for the Miras in Sgr~I.
The long dashed area shows the contribution in the galactic bulge
and the dashed line is the relation from Glass et al., shifted 2\mag35
(i.e. 24~kpc if the galactic centre is at 8~kpc)}
\endfig
\hyphenation{Azzopardi}
The Mira and semiregular variables in our sample of possible members
of the dwarf galaxy are best looked at in a period-luminosity diagram
(Fig.~2). Whitelock et al. (1991) demonstrated that the 
period-luminosity relation is
independent of the metallicity of the Miras. This might also
apply to the semiregulars. In Paper~II it is 
argued that they are the short period extension to this relation.
In Fig.~2 we plotted the variables in Table~1, together with the
galactic contribution in PG3. We refer to Paper~II 
for details about the galactic contribution.
We also show the PK$_0$-relation of Glass et al. (1995) shifted 
2\mag35, i.e. to a distance of 24~kpc. 
At this distance the stars are about 4 kpc out of the galactic 
plane, where one expects to find mainly old metal-poor stars,
from the metal-poor thick disc and the halo (Ng et al. 1996). 
In these populations one does not expect 
to find long period variable stars, because they
have not been found in the old metal-poor globular clusters.
Therefore, the semiregular variables and Miras cannot be of galactic origin 
and should belong to a considerable younger population. 
But a dwarf galaxy, which has interaction with our Galaxy, 
can contain younger populations.
This demonstrates that the long period variables 
could indeed be located in the dwarf galaxy, except for 
variables \#192 and \#1128 which might have a galactic origin.
Additional photometry is required to determine this.
\par
One of the variables (\#283)
is a carbon star, see Fig.~3.
Details about the spectroscopic observations and the spectral classification
of the Miras and 
the semiregular variables will be given in a forthcoming paper 
(in preparation).
This star appears to be comparable with the carbon
stars found in other dwarf galaxies (Aaronson et al. 1983,
Azzopardi et al. 1985 \& 1986) and  
some of the carbon stars found by Azzopardi et al. (1991).
Unfortunately, the spectrum of variable \#283 did not include the 
sodium D doublet and it is not clear if this star is comparable 
with those found in the `bulge' or the low-metallicity objects 
in the SMC and the dwarf galaxies.
The `bulge' carbon stars have been a mystery (Lequeux 1990,
Tyson \& Rich 1991, Westerlund et al. 1991), because they are about
2\mag5 in bolometric luminosity too faint to be regarded 
as genuine AGB stars, if located inside the metal-rich bulge. 
But if some or all of them  
are located in the dwarf galaxy, just like variable \#283,
there is no need for a metal-rich origin.
They are in that case 
just ordinary metal-poor to intermediate metallicity 
carbon stars.
\par
\begfigps 5.5cm
\psfig{file=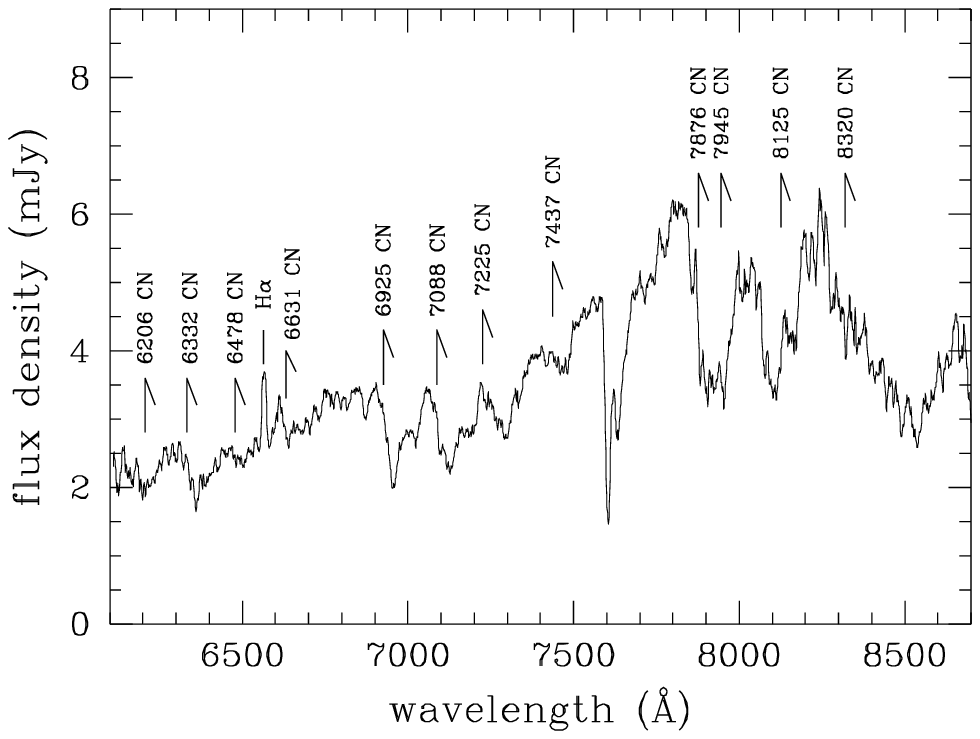,height=5.5cm,width=8.6cm}
\figure{3}{Medium resolution spectrum smoothed to
2.6\,\AA/pixel for variable star \#283 (C 3,2), which has a period 
\hbox{P\muspc=\muspc137.39} days (Wesselink 1987)}
\endfig
\noindent
Membership of the dwarf galaxy has important consequences. 
It implies that the carbon stars are at least younger than 
approximately 4~Gyr (Marigo et al. 1996 and references cited therein). 
It indicates that 
at least 2 major epochs of star formation occurred in the 
dwarf galaxy, just like the recurrent star formation epochs 
in the Carina dwarf spheroidal (Schmecker-Hane et al. 1996).
It possibly traces the tidal star formation tail due to its passage through 
our Galaxy. This tail extends at least from $b\!=\!-2\fdg4$
to $b\!=\!-9\fdg9$, which is far more larger than previously
thought. This tail most likely indicates that the Sagittarius
dwarf galaxy has passed the galactic plane not so long ago.
It would explain the rather blue colours and 
might also explain the large velocity dispersion of the carbon stars 
obtained by Tyson \& Rich (1991).
\par
Alksnis (1990) showed that the majority of the carbon stars are 
SRa-type long period variables. A detailed monitoring of the 
Azzopardi et al. (1991) carbon stars is required to determine 
if they are variable and to determine their periods. Together 
with the period-luminosity relation their membership to the
dwarf galaxy can be secured.
This can be compared with the properties of the carbon stars studied 
by Whitelock et al. (1996). Their study indicates the presence of two
different groups. One group with  
\hbox{(J--K)$_0$\muspc$<$\muspc1\mag3} is comparable with the
carbon stars from Azzopardi et al. (1991), while another group
has significantly redder colours \hbox{(J--K)$_0$\muspc$>$\muspc1\mag7}.
The two groups could be an indication for two different star formation
epochs from recent passages through the galactic plane.
\par
AGB stars, like the carbon stars, are the progenitors of planetary
nebulae (PN). Along the trail of the Azzopardi et al. (1991) 
carbon stars one would expect to find long period variables and
PNs. The question arises if the PNs found 
at low galactic latitude with velocities near to that of the 
Sagittarius dwarf galaxy (Zijlstra \& Walsh 1996) should be
considered as true bulge members? Membership of the dwarf galaxy 
would support the proposition that some or all of 
the Azzopardi et al. 
carbon stars do indeed trace the tidal tail.

\acknow{The research of MS is supported by a grant from the
Austrian Science Fund under project
number P9638--AST and S7308. 
YKN is supported by HCM grant CHRX-CT94-0627
from the European Community.}

\begref{References}
\ref Aaronson M., Olsewski E.W., Hodge P.W., 1983, ApJ 267, 271
\ref Alard C., 1996, ApJ 458, L17
\ref Alksnis A., 1990, in proceedings 
``{\it From Miras to planetary nebulae: Which path for 
stellar evolution?}'', Montpellier (France), Sept. 4-7, 1989,
M.O. Menessier and A. Omont (eds.), Editions Fronti\`eres, p279
\ref Azzopardi M., Lequeux J., Westerlund B.E., 1985, A\&A 144, 388
\ref Azzopardi M., Lequeux J., Westerlund B.E., 1986, A\&A 161, 232
\ref Azzopardi M., Lequeux J., Rebeirot E., Westerlund B.E.,
1991, A\&AS 88, 265
\ref Blommaert J.A.D.L., 1992, Ph.D. thesis, 
Leiden University, the Netherlands
\ref Glass I.S., Whitelock P.A., Catchpole R.M., Feast M.W.,
1995, MNRAS 273, 383
\ref Ibata R., Gilmore G., Irwin M.J., 1994, Nature 370, 194
\ref Lequeux J., 1990, in proceedings 
``{\it From Miras to planetary nebulae: Which path for 
stellar evolution?}'', Montpellier (France), Sept. 4-7, 1989,
M.O. Menessier and A. Omont (eds.), Editions Fronti\`eres, p273
\ref Mateo M., Mirabel N., Udalski A., et al., 1996, ApJ 458, L13
\ref Marigo P., Bressan A., Chiosi C., 1996, A\&A {\it in press}
\ref Mateo M., Kubiak M., Szyma\'nski M. et al., 1995, AJ 110, 1141
\ref Mateo M., Mirabel N., Udalski A., et al., 1996, ApJ 458, L13
\ref Ng Y.K., 1994, 
Ph.D. thesis, Leiden University, the Netherlands
\ref Ng Y.K., Bertelli G., Bressan A., 
Chiosi C., Lub J., 1995, A\&A 295, 655 {(erratum A\&A 301, 318)}
\ref Ng Y.K., Bertelli G., Chiosi C., Bressan A., 1996, A\&A {\it submitted}
\ref Oort J.H., Plaut L., 1975, A\&A 41, 71
\ref Plaut L., 1971, A\&AS 4, 75
\ref Schmecker-Hane T.A., Stetson P.B., Hesser J.E., VandenBerg D.A., 
1996, Proceedings `From Stars to Galaxies', ASP Conference Series 
Vol. 98, C. Leitherer, U. Fritze-von Alvensleben and J. Huchra (eds.), 328
\ref Schultheis M., Ng Y.K., Hron J., Kerschbaum,\ F., 1996, 
A\&A {\it submitted} (Paper~II)
\ref Tyson N.D., Rich R.M., 1991, ApJ 367, 547
\hyphenation{Nij-me-gen}
\ref Wesselink Th.J.H., 1987, Ph.D. thesis, Catholic University
Nijmegen, the Netherlands
\ref Westerlund B.E., Lequeux J., Azzopardi M., Rebeirot E.,
1991, A\&A 244, 367
\ref Whitelock P.A., Feast M.W., Catchpole R.M., 1991, MNRAS 248, 276
\ref Whitelock P.A., Irwin M., Catchpole R.A., 1996, New Astronomy 1, 57
\ref Zijlstra A.A., Walsh J.R., 1996, A\&A 312, L21
\endref
\bye